\documentclass[twocolumn]{aa}
\usepackage{graphicx}
\usepackage{txfonts}
\def\ref{\hfil\goodbreak\vskip 0.15 in\noindent\hangindent 20pt {}}
\begin{document}
\title{Dynamics and Heating of the Magnetic Network on
the Sun} 
\subtitle{Efficiency of mode transformation}
\author{S. S. Hasan 
\inst{1} 
\and
P. Ulmschneider \inst{2} 
}
\offprints{S. S. Hasan}
\institute{Indian Institute of Astrophysics, Bangalore-560034, India\\
  \email{hasan@iiap.res.in}
\and
Institute f\"ur Theoretische Astrophysik, Tiergartenstr. 15, 69121 Heidelberg, Germany\\
   \email{ulm@ita.uni-heidelberg.de}
}
%\maketitle
%\begin{sloppypar}
\date{}
\abstract{
We aim to identify the physical processes which occur in the magnetic
network of the chromosphere and which contribute to its dynamics and
heating.  Specifically, we study the propagation of transverse (kink)
MHD waves which are impulsively excited in flux tubes through
footpoint motions.  When these waves travel upwards, they get
partially converted to longitudinal waves through nonlinear effects
(mode coupling). By solving the nonlinear, time-dependent MHD
equations we find that significant longitudinal wave generation
occurs in the photosphere typically for Mach numbers as low as 0.2 and
that the onset of shock formation occurs at heights of about 600 km
above the photospheric base. We also investigate the compressional
heating due to longitudinal waves and  the efficiency of mode
coupling for various values of the plasma $\beta$, that parameterises
the magnetic field strength in the network.  We find that this
efficiency is maximum for field strengths corresponding to
$\beta\approx 0.2$, when the kink and tube wave speeds are almost
identical. This can have interesting observational
implications. Furthermore, we find that even when the two speeds are
different, once shock formation occurs, the longitudinal and
transverse shocks exhibit strong mode coupling.  
}
\authorrunning{S.S.Hasan \& P. Ulmschneider}
\titlerunning{Dynamics and heating of the magnetic network}
\maketitle
\section{Introduction}
The solar chromosphere plays an important role as the lower boundary
of the heliosphere, as the source of the solar XUV radiation which
affects the ionisation state of the upper terrestrial atmosphere as
well as for the origin of the solar wind and the generation of coronal
mass ejections.  It is thus of great interest to understand the
chromosphere and, in particular, the state of the gas in its upper
layers.  In the quiet chromosphere one needs to distinguish between
the magnetic network and internetwork regions in the cell interior. The
former occurs on the boundary of several supergranulation cells which
are the sites of the ubiquitous Network Bright Points (NBPs) in which
strong magnetic fields are organized in magnetic flux tubes. On the
other hand, magnetic fields are weak and dynamically unimportant in
the internetwork.

The focus of this investigation is to examine wave propagation in
magnetic elements of the network. Several observations have revealed
that network oscillations have periods that are much longer than those
in internetwork regions (Lites, Rutten \& Kalkofen 1993; Judge,
Carlsson \& Wilhelm 1997; Curdt \& Heinzel 1998; Cauzzi, Falchi \&
Falciani 2000). A careful analysis of NBPs has revealed that these
oscillations possess significant power in the 1-4 mHz (Banerjee et
al. 2001; McAteer et al. 2002, 2003; Bloomfield et al. 2004), along
with evidence which suggests the existence of multiple peaks.

A theoretical understanding of the waves observed in the network is
far from complete. An idealized scenario is to treat the network as a
conglomeration of thin vertical magnetic flux tubes fanning out with
height. It is well known that such flux tubes support a variety of
wave modes: the sausage or longitudinal mode, the kink or transverse
mode, and the torsional Alfv\'en mode (Spruit 1982; Roberts
\& Ulmschneider 1997). This approach was adopted by  Musielak et al. (1989,
1995), Huang, Musielak \& Ulmschneider (1995) and Ulmschneider \&
Musielak (1998), Musielak and Ulmschneider (2001,2002,2003a,b), Noble,
Musielak \& Ulmschneider (2003) to
examine MHD wave generation in flux tubes through an interaction with
turbulent motions in the convection zone. In these studies, the
turbulence was characterised using a modified Kolmogorov spectrum. An
alternative model, based on G-band observations of NBPs motions
(Muller 1983, 1985), investigated the excitation of MHD kink waves in
flux tubes through impulsive motions of their footpoints (Choudhuri,
Auffret \& Priest 1993). 

The latter hypothesis was investigated in greater
detail by Hasan \& Kalkofen (1999) who examined the generation of
transverse and longitudinal waves in a flux tube through buffeting by
granular motions in the magnetic network. It was found that the
generic response of the flux tube to a single granular impact is the
same for both transverse and longitudinal waves: the buffeting action
excites a pulse that propagates along the flux tube with the kink or
longitudinal tube speed. For strong magnetic fields, most of the
energy goes into transverse waves, and only a much smaller fraction
into longitudinal waves. After the passage of the pulse, the
atmosphere gradually relaxes to a state in which it oscillates at the
cutoff period of the mode. The period observed in the magnetic network
was interpreted as the cutoff period of transverse waves, which leads
naturally to an oscillation at this period (typically in the 7-minute
range) as proposed by Kalkofen (1997).

As a continuation of the above investigation, Hasan, Kalkofen \& van
Ballegooijen (2000) modeled the excitation of waves in the network due
to the observed motions of G-band bright points, which were taken as a
proxy for footpoint motions of flux tubes. For a typical magnetic
element in the network they predicted that the injection of energy
into the chromosphere takes place in brief and intermittent bursts,
lasting typically 30 s, separated by long periods with low energy
flux; this implies a high intermittency in chromospheric emission,
which may be incompatible with observations.  They concluded that
there must be other high-frequency motions (periods 5-50 s) which
cannot be detected as proper motions of G-band bright points. Adding
such high-frequency motions to their simulations they obtained much
better agreement with the persistent emission observed from the
magnetic network. They speculated that the high-frequency motions
could be due to turbulence in intergranular lanes. This idea, however, requires
further investigation.

The above studies were based on a linear approximation in which the
longitudinal and transverse waves are decoupled.  However, it is well
known that the velocity amplitude $v(z)$ for the two modes increases
with height $z$ (for an isothermal atmosphere $v \propto \exp [z/4H]$,
where $H$ is the pressure scale height), so that the motions are
likely to become supersonic higher up in the atmosphere. At such
heights, nonlinear effects become important, leading to a coupling
between the transverse and longitudinal modes.  Some progress on this
question has been made in one dimension (1-D) using the nonlinear
equations for a thin flux tube (Ulmschneider, Z\"ahringer \& Musielak
1991; Huang, Musielak \& Ulmschneider 1995), Zhugzhda, Bromm \&
Ulmschneider (1995) and more recently by Hasan et al. (2003, hereafter
Paper I). 

In Paper I, the nature of mode coupling between the
transverse and longitudinal modes in the magnetic network was
investigated in some detail. They modelled the excitation of kink
oscillations through impulsive footpoint motions of flux tubes and
found that when the transverse velocities are significantly less than
the kink wave speed (the linear regime), there is essentially no
excitation of longitudinal waves.  However, at heights where the Mach
number of the kink oscillations is around 0.3, longitudinal modes
begin to be excited.  Longitudinal wave generation becomes most
efficient at Mach numbers around unity, leading to the modes having
comparable amplitudes.  A comparison of the results with the exact
linear solution for transverse waves enabled them to locate the
regions in the atmosphere where nonlinear effects are important.

The main aim of the present study is to extend the calculations of
Paper I to examine the efficiency of mode conversion for different
values of $\beta$, which essentially is a measure of the magnetic
field strength in the magnetic network. The motivation for such a
study stems from trying to understand whether mode conversion would be
efficient over a wide range of $\beta$ values or whether it holds only
for a narrow range of $\beta$. In our scenario, the heating of the
network elements is assumed to occur through the dissipation of
longitudinal waves, which are generated from transverse waves through
mode conversion. Therefore, our investigation could be useful in
placing constraints on the field strengths in the network where
chromospheric heating occurs through the above process.

The organization of this paper is as follows: in Sect. 2 we discuss
the excitation of MHD waves by footpoint motions in a vertical
``thin'' flux tube.  along with the initial equilibrium model and the
method of solution. The results of our calculation are presented in
Sect. 3 followed by a discussion and summary in Sect. 4, where we
also point out some observational implications of our investigation.

\section{Wave Excitation Due to Footpoint Motion}
\def\p{\perp}   \def\V{{\cal  V}} \def\e{{\rm e}}  
\def\e#1{{\rm e}^{#1}}  \def\ex#1{{e}^{#1}}\def\ei#1{{\rm e}^{{\rm i}#1}}

In this section we study wave propagation in a tube over a height
range where the thin-flux tube approximation (TFA) can be applied. The
validity of this approximation to flux tubes in the solar atmosphere
is discussed in some detail in Paper I. By comparing the structure of
a thin flux tube with that calculated using a 2-D model, it was shown
that the TFA can be applied up to heights of around 1000 km above the
photospheric base. At larger heights, the TFA deteriorates owing to
the expansion of the flux tube near the ``canopy''.

\subsection{Initial flux tube model}
Similar to Paper I, let us consider a flux tube extending vertically
through the photosphere and chromosphere of the Sun.  We assume that
the flux tube is ``thin'' and initially in hydrostatic equilibrium and
isothermal, with the same temperature as the external medium, which we
take to be 6000~K. We consider a tube with a radius of 50 km at
$z=0$. The radius increases with $z$ as $\exp(z/4H)$ in an isothermal
atmosphere for a thin flux tube.

\begin{figure*}
%\centerline{ \psfig{figure=vb1425.epsi,height=11cm} }
\centering
\includegraphics{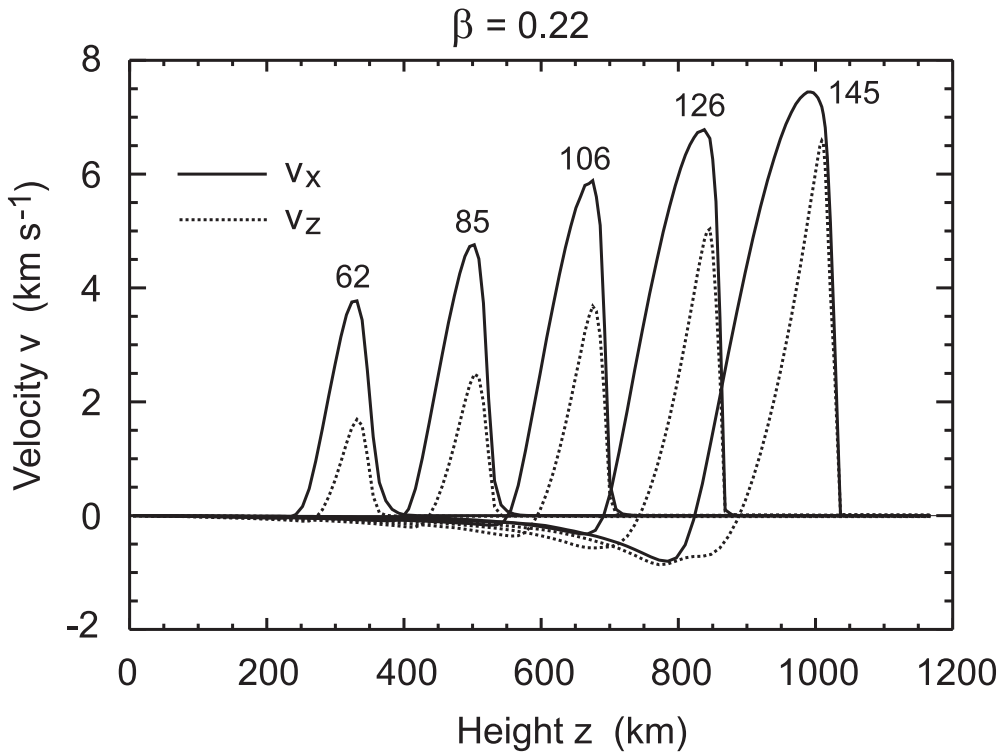}
\caption{Nonlinear coupling of transverse and longitudinal waves in a
flux tube: Transverse velocity $v_x$ ({\it solid curves}) and
longitudinal velocity $v_z$ ({\it dashed curves}) as functions of
height $z$ at various times, for $\beta=0.22$, $t_0 = 20$~s, $\tau =
5$~s, $v_0=2.5$ $\rm km~s^{-1}$. The numbers adjacent to the curves
denote the time in seconds. }
\end{figure*}

\subsection{Method of solution}
The basic equations for adiabatic longitudinal-transverse MHD waves in
a thin flux tube consist of a set of coupled differential equations
(see Ulmschneider, Z\"ahringer and Musielak 1991 for details) which
are solved numerically using the method of characteristics.  Our method 
is based on a
modified version of the code developed by Ulmschneider, Z\"ahringer \&
Musielak (1991) that treats shocks (following 
Zhugzhda, Bromm \& Ulmschneider 1995) as well as the back reaction of the
external fluid on the tube (Osin, Volin \& Ulmschneider 1999).  The code is capable
of resolving shocks in the chromosphere. However, we have made an
important modification to this code by incorporating the transmission of
strong shocks through the upper boundary. The computational domain in
the vertical direction has an equidistant grid of size about 9 km. The
Courant condition is used to select the time step to advance the
equations in time.
  
\subsection{Boundary conditions}
At the lower boundary, taken at $z=0$, we assume that 
the flux tube has a transverse footpoint motion consisting of a single impulse  
of the form:
\begin{equation}
v_{x}(0,t)=v_0\ex{-[(t-t_0)/\tau]^2}
\label{eq:a1}
\end{equation}
where $v_0$ is the specified velocity amplitude, $t_0$ denotes the time
when the motions have maximum amplitude and $\tau$ is the time constant
of the impulse. The longitudinal component of the velocity at the base is
assumed to be zero.  In the present calculations we take $v_0=2.5$
km~s$^{-1}$, $t_0=20$~s and $\tau=5$~s. 

\begin{figure*}
%*\centerline{ \psfig{figure=veltemp.epsi,height=20cm} }
\includegraphics{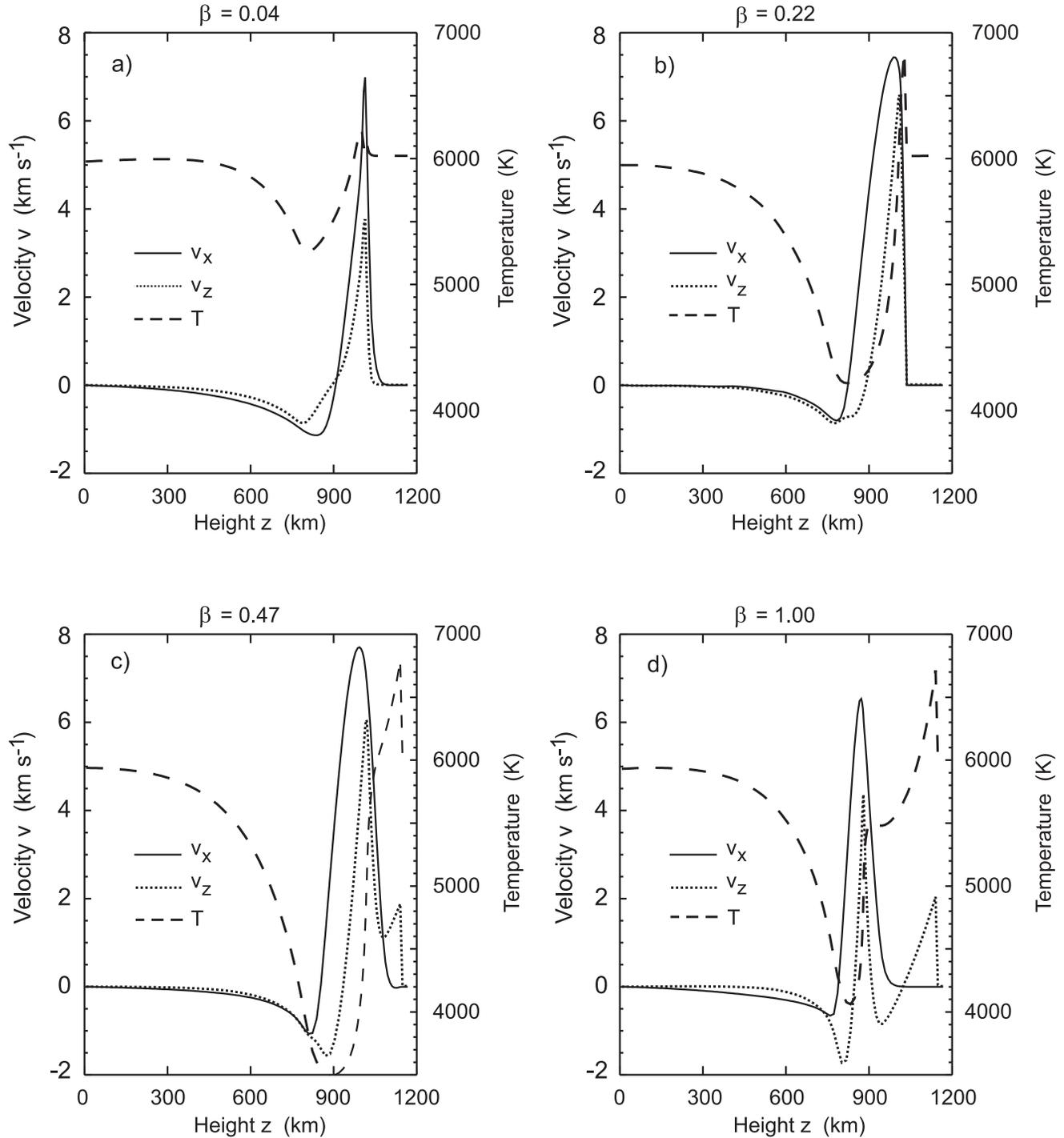}
\centering
\caption{
Snapshots of the transverse velocity $v_x$ ({\it solid curves}) and
longitudinal velocity $v_z$ ({\it dotted curves}) and the temperature
$T$ (dashed curves) as functions of height $z$ for (a) $\beta=0.04$,
(b) $\beta=0.22$, (a) $\beta=0.47$, and (d) $\beta=1.00$. }
\end{figure*}

At the upper boundary of the computational domain (taken at $z=1200$ km) we
use transmitting boundary conditions, following Ulmschneider et
al. (1977), and assume that the velocity amplitude remains constant
along the outward-propagating characteristics. We use the characteristic
equations to self-consistently determine physical quantities
at the boundary.

\begin{figure*}
%\centerline{ \psfig{figure=vcompbeta.epsi,height=11cm} }
\includegraphics{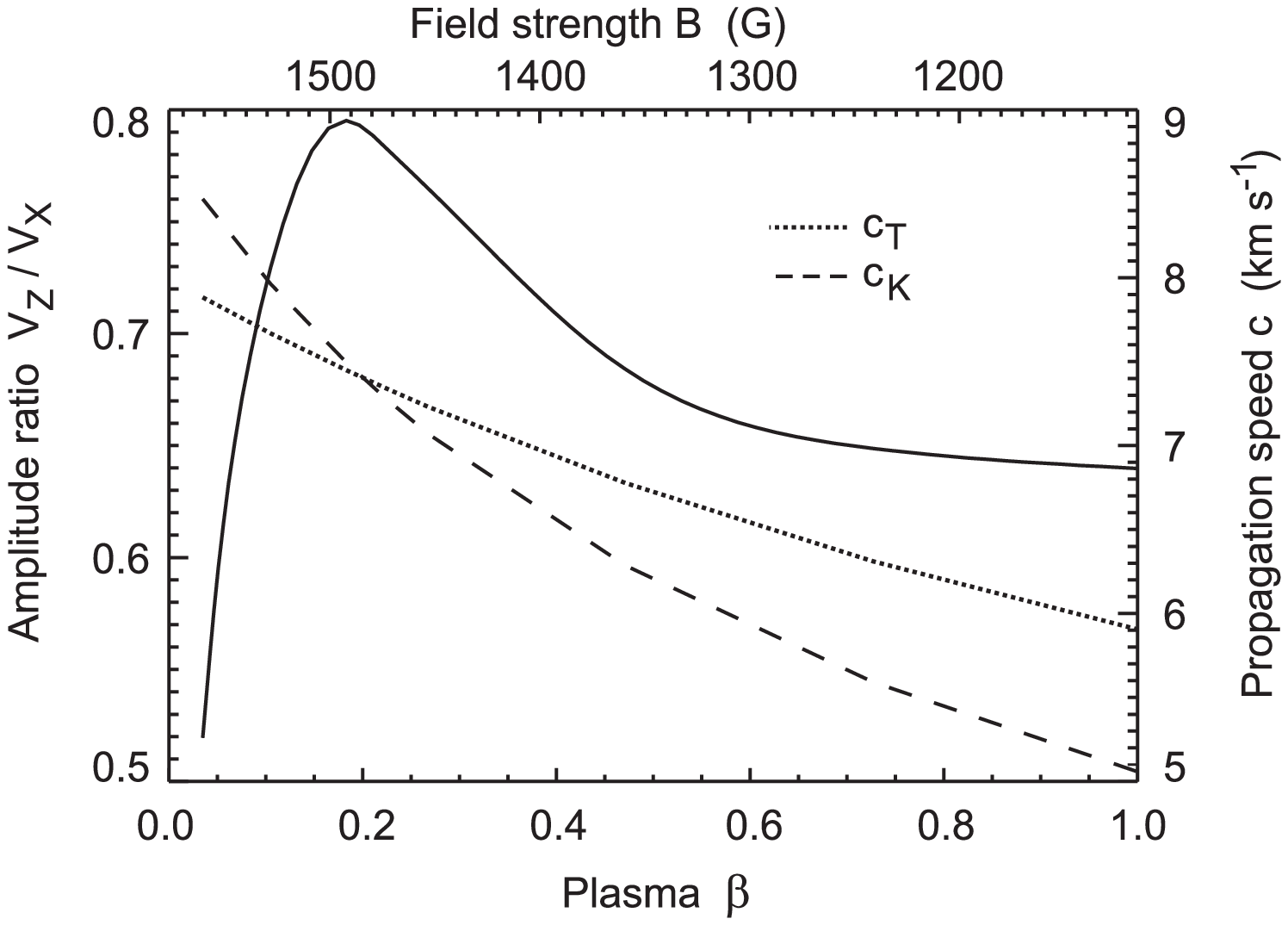}
\centering
\caption{Efficiency of nonlinear coupling between transverse and longitudinal waves in a
flux tube: $v_z/v_x$ ({solid curves}) at $z=900$~km, $c_K$ (dashed
curve) and $c_T$ (dotted curve) as functions of $\beta$. The field
strength $B$ at $z=0$ is given by the top axis.}
\end{figure*}

\section{Results}
The initial equilibrium model is perturbed with a transverse motion at
$z=0$ in the form of an impulse with a velocity given by equation (1).
This impulse generates a transverse wave that propagates upwards with
the kink wave tube speed.  The resulting motion in the tube as a
function of height and time follows from the time-dependent MHD
equations for a thin flux tube.

Figure 1 shows the variation of the transverse $v_x$ (solid lines) and
longitudinal $v_z$ (dashed lines) components of the velocity as a
function of height $z$ at various epochs of time $t$ for a magnetic
field strength of $1425$~G at $z=0$, that corresponds to a plasma
$\beta=0.22$ (which by assumption remains constant with height for a
thin flux tube). In the present case, the kink and longitudinal tube
speeds are approximately the same with a value of about 7.3 km
s$^{-1}$. The numbers next to the curves denote the time $t$ (in s).
We find that low in the atmosphere, where the transverse velocity
amplitude is small (compared to the kink wave speed $c_K$), the
longitudinal component of the velocity is negligible. As the initial
pulse propagates upwards, the transverse velocity amplitude
increases. Due to nonlinear effects, longitudinal motions are
generated. We find that even at fairly low heights ($z\approx 300$~km),
the latter are significant for a Mach number $M=v_x/c_K$ as small as
0.2. The longitudinal pulse propagates at the tube speed $c_T$, which
in the present case is almost identical with $c_K$.  The efficiency of
the nonlinear coupling increases with the amplitude of the transverse
motions. When $v_x\approx c_K$, the amplitudes in the transverse and
longitudinal components become comparable. The longitudinal motions,
being compressive, steepen with height and eventually form shocks. A
shock is first formed at a height of about 600 km. This is a
kink-longitudinal shock as discussed in detail by Zhugzhda, Bromm \&
Ulmschneider (1995), which travels upward with the common propagation
kink/tube speed. We expect the compression associated with the
longitudinal shock to produce heating as we shall see in the next
figure.

Figures 2(a)-(d) depict the height variation of the horizontal and
vertical components of the velocity (solid and dotted lines
respectively) and the temperature $T$ (dashed line) at a certain
instant of time for various values of $\beta$. In figure 2(a), we find
that due to mode coupling, the transverse and longitudinal shocks
travel approximately together, despite the fact that the phase speeds
for the two modes are different ($c_K=7.9$ km s$^{-1}$ and $c_T=8.5$
km s$^{-1}$). In figure 2(b), we find that the location of the shocks
occurs at identical heights, which is not surprising since in this
case $c_K=7.9\approx c_T$=7.3 km s$^{-1}$. As we increase $\beta$ further, so
that $c_K < c_T$, an interesting phenomenon occurs. The longitudinal
shock splits into two shocks, one occurring at about the same location
as the kink shock, but another one of smaller magnitude occurring at a
greater height. The latter travels with the speed $c_T$. 

We now consider the effect of the waves on the temperature structure
in the flux tube. Through nonlinear interactions, longitudinal waves
are generated from purely transverse wave excitation. Being
compressive, longitudinal waves heat the atmosphere. The dashed curves
in figures 2(a)-(d) show the temperature enhancement due to the
compressive longitudinal waves. In figures 2(a) and 2(b) corresponding
to strong magnetic fields for which $c_K > c_T$, the amplitude of the
temperature pulse increases with the longitudinal velocity
amplitude. However, the situation is more complicated for weaker
fields ($c_K < c_T$).From figures 2(c) and 2(d), we find that the
shock heating does not occur at the common transverse-longitudinal
wave front but rather at a larger height associated with a
longitudinal shock.

Finally, we examine the efficiency of the mode coupling as a function
of $\beta$, which parameterises the magnetic field strength in the
flux tube. The mode coupling efficiency at a particular height can be
measured in terms of the ratio $v_z/v_x$, since the longitudinal
motions are essentially generated from the transverse motions through
nonlinear interactions. Figure 3 shows the variation of $v_z/v_x$ with
$\beta$ at $z=900$~km. The top scale shows the corresponding values of
the magnetic field strength at $z=0$. The dashed and dotted lines
denote the kink and tube speeds respectively (with respect to the
scale on the right). For low values of $\beta$ (i.e. for which $c_K >
c_T$) the efficiency increases rapidly with $\beta$ reaching a maximum
at $\beta\approx 0.2$ i.e., when the kink and tube speeds are
equal. When $c_T > c_K$, the efficiency drops with increasing
$\beta$. At large values of $\beta$, $v_z/v_x$ appears to level off
to a limiting value.

\section{Discussion and Summary}
The results presented in the previous section have highlighted several
interesting features connected with wave propagation in magnetic flux
tubes. Similar to Paper I, we found that through the forcing action of
purely transverse motions at the base of a flux tube, longitudinal
motions are generated higher up in the tube through mode coupling. The
new refinements in our present code, allow us to accurately follow
shocks as well as to examine the coupling of the modes to large
heights in the atmosphere.

\subsection {Linear mode coupling at similar propagation speeds}
The focus of this work has been on examining the effect of
$\beta$ on the mode coupling between the transverse and longitudinal
waves in the solar atmosphere. It is, of course well known, that a
magnetized atmosphere exhibits coupled slow and fast modes which for a
low $\beta$ plasma are akin to acoustic and magnetic modes
respectively. Interesting effects occur when $\beta$ varies with
height in the atmosphere. Bogdan et al. (2003) point out that the
$\beta=1$ layer, which they identify with the magnetic canopy, is
crucial in understanding the nature of MHD wave propagation in the
solar chromosphere. In this layer, the fast and slow MHD waves
experience strong coupling which leads to ``mode mixing''(e.g. Campos
1987 and Alicki et al. 1994). It should be noted that this coupling is
essentially a linear effect which is due to the change in the relative
strengths of the sound and Alfv\'en speeds in the atmosphere.  We
should like to clarify that in the framework of the thin flux tube
approximation, however, there is no linear coupling between the
transverse and longitudinal modes. The coupling that we discuss in
this paper arises solely due to nonlinear effects. The physical origin
of this phenomenon has been discussed by Ulmschneider, Z\"ahringer
\& Musielak (1991) (see also Ulmschneider 2003). 

\subsection {Nonlinear mode coupling of waves and shocks}
Let us now consider the non-linear mode coupling between the
transverse and longitudinal waves. In this context we would like to
distinguish between two effects: (a) the generation of longitudinal
waves through purely transverse displacements at the base of the flux
tube and, (b) the coupling between longitudinal and kink shocks. The
first effect has already been discussed in Paper I as well as by
Ulmschneider, Z\"ahringer \& Musielak (1991) or Ulmschneider (2003)
and no further elaboration is necessary, other than the fact that its
efficiency is highest when the tube and kink waves speeds become
identical.

On the other hand, even when the two speeds are different, so that the
nonlinear interaction at lower wave amplitude (see figure 2a) is less
efficient, we find that as soon as the shock forms, the transverse
shock and the longitudinal shock stay at the same height and propagate
with the same common speed, different from both the longitudinal and
transverse wave speeds. This new property of shock waves suggests a
particularly large and unusual amount of mode coupling. It 
has already been found in earlier calculations (Zhugzhda, Bromm \&
Ulmschneider 1995; Hollweg, Jackson \& Hollweg 1982).  The fact that this
feature seems to be a fundamental property of shock waves and is not
an artifact of our wave computation can be seen by noting that it
also occurs in other time-dependent MHD wave computations using
different wave modes and other mathematical methods.  We should point out that our
computations are based on the method of characteristics where shocks are
treated as discontinuities at which the two sides are connected with
the help of the Hugoniot relations whereas the treatment of Hollweg et
al. use a finite difference method where the treatment of shocks is
not different from that of other interior points.

Figures 3 and 4 by Hollweg, Jackson \& Galloway (1982) show the
propagation of torsional wave pulses in a solar magnetic flux
tube. These authors examined the time development of wave pulses of 90
s duration with different initial torsional velocity amplitudes. It is
seen that these pulses propagate towards greater height and eventually
develop strong torsional shocks (called switch-on shocks). Mode
coupling also generates longitudinal waves.  A comparison of the shock
position of the torsional and longitudinal shocks at the various times
shows that the shocks are always at the same height and propagate at
the same common speed, exactly like the behaviour which we found in
Figure 2 for our coupled longitudinal-transverse wave system.

\subsection{Mode coupling and heating}
Let us now turn to the question of wave heating associated with the
motions in the flux tube.  As can be seen from figures 2(a)-(d),
heating takes place in the tube due to the occurrence of shocks. For
very small values of $\beta$, when the kink wave speed $c_K > c_T$,
we find that shock heating is minimal. The largest amount of shock
heating at the common kink-longitudinal shock occurs when the
transverse and longitudinal wave speeds are approximately identical,
and where $v_z\approx v_x$.  For weaker fields ($c_K < c_T$), the
heating at the common shock front again becomes negligible. However,
in this case, the presence of the additional purely longitudinal shock
produces heating ahead of the kink-longitudinal shock. This is another
new feature to emerge from the present calculations. We should clarify
that although we use an adiabatic energy equation that assumes entropy
conservation in the flux tube, we use the Rankine-Hugoniot relations
at the shock front to connect the pre and post shock states. It is the
entropy jump at the shock front which is responsible for the heating
discussed above.

It would be premature for us to estimate, on the basis of the
calculations in this paper, the wave energy flux entering the corona
or even the upper chromosphere. This is mainly on account of two
assumptions that have been made by us. The first one is our choice of
an isothermal atmosphere which implies that the pressure scale height
and the sound speed are constant with height $z$. For the height range
considered by us, the sound speed variation is less than 10\% and is
unlikely to affect the results in a fundamental way. 

The second assumption which we have made that is more serious is the
thin flux tube approximation, whose validity was carefully examined in
Paper I. By comparing the exact magnetostatic solution for an
axisymmetric flux tube, with that obtained using the thin flux tube
approximation, it was shown that the latter is reasonable up to
heights of about 1 Mm above the photosphere for tubes with a filling
factor less than 1\%. on the solar surface. At larger heights even a
sufficiently thin flux tube expands rapidly and comes in contact with
neighbouring tubes, making the problem intractable within the present
formalism. In such a situation, where further spreading of the tube is
inhibited, $\beta$ is no longer constant but typically increases with
height and the shape of the tube in the upper atmosphere resembles a
``wine glass''. However, in our investigation we have used an upper
boundary at a location where the exponentially diverging tube
behaviour approximates the true solution reasonably well.

\subsection{Mode coupling in the wine glass geometry}
A major difficulty of using a modified thin flux tube with a
``wine glass'' in the upper chromosphere stems from the difficulty of
including the effects of waves from different flux tubes interfering
with one another at such locations. Clearly, one needs to resort to
multidimensional calculations to treat the problem
accurately. Recently, some progress has been made in this direction
(Rosenthal et al. 2002 and Bogdan et al. 2003) who have carried out
numerical simulations of wave propagation in 2-D, assuming that the
displacements lie in a plane perpendicular to the invariant direction,
for magnetic geometries representative of the network and
inter-network regions on the Sun.  The velocity amplitudes in the
above simulations are sufficiently small so that nonlinear effects are
relatively unimportant compared to those associated with the field
geometry.  Furthermore, these waves are different from the kink and
sausage modes considered by us, for which three-dimensional
simulations would be required, that have not yet been carried
out. Nevertheless, the above work represents an important step which
should be followed up by more refined calculations to allow a
realistic comparison with observations.

\subsection{Observational implications}
It is interesting to speculate on the observational implications of
our calculations. Observationally, it is hard to unambiguously test
whether there is evidence of mode coupling in the network based on the
scenario presented by us. McAteer et al. (2003) and Bloomfield et
al.(2004) examined several NBPs and inferred the possibility of mode
coupling between transverse and longitudinal modes in about a third of
them. This inference was based on cross-correlating power in different
lines to look for signatures of transverse and longitudinal modes at
different heights in the atmosphere. In the lower atmosphere, the
modes are dominantly transverse with a frequency in the range 1.2-1.6
mHz that can be identified with the kink cutoff frequency. Through
mode coupling, longitudinal waves are generated at double the
frequency (e.g. Musielak and Ulmschneider 2003b).  On the basis
of the present theoretical study, it is difficult to predict the
dominant frequencies for the transverse and longitudinal waves in NBPs
other than the presence of their respective cutoff frequencies (see
Paper I) if the footpoint excitation occurs in the form of a single
pulse. However, for a more realistic excitation mechanism, which on
observationally grounds is likely to be stochastic in nature, further
work is required before any predictions on mode frequencies can be made. We
hope to examine this problem in greater detail in future work. 

The present calculations would suggest that an investigation of a
large sample of NBPs would tend to show weak mode coupling for very
large field strengths (at the photospheric base) greater than about
1500 G. For weaker field strengths, the mode coupling efficiency
would be correlated with the field strength at $z=0$. This may be
consistent with why McAteer et al. (2003) and Bloomfield et al. (2004) did
not find evidence of mode coupling in all the NBPs they examined. We
expect that future observations will shed more light on this question.

In summary, the purpose of the present investigation was to examine
the effect of the magnetic field strength (parameterised by $\beta$) on
the dynamics, heating and efficiency of mode coupling in flux tubes
that are typical of the magnetic network on the Sun. We modelled the
excitation of kink oscillations through impulsive footpoint
motions. Through nonlinear mode coupling, longitudinal
motions are generated. For typical parameters in the network, we found
that significant longitudinal wave generation occurs at heights as low
as 300 km in the atmosphere and that shock formation occurs at about
$z=600$ km, which can lead to heating of the chromosphere. The
efficiency of mode coupling is maximum when the kink and tube waves
speeds are almost identical, which occurs for $\beta\approx 0.2$. We
also pointed out that the present calculations can have interesting
observational implications. In future work we hope to overcome the
limitations of the thin flux tube approximation to examine more
thoroughly chromospheric heating and also the transmission of wave
energy into the corona. 

\begin{acknowledgements}
We have benefited from the valuable comments and suggestions of the
referee Z. Musielak. SSH gratefully acknowledges support from the
Deutsche Forschungs Gemeinschaft to visit the Institut f\"ur
Theoretische Astrophysik, University of Heidelberg for carrying out
the work reported in this paper. SSH is thankful to K. Sivaraman for
helpful discussions.
  
\end{acknowledgements}

%\end{sloppypar}

\begin{thebibliography}{qqq}
\bibitem{} Alicki, R., Musielak, E. Z., Sikorski, J. \& Makowiec, D. 1994, 
            ApJ, 427, 919
\bibitem{} Banerjee, D., O'Shea, E., Doyle, J. G., \& Goosens, M. 
  2001, A\& A, 357, 1093
\bibitem{} Bloomfield, D. S., McAteer, R. T. J., Mathioudakis, M., 
Williams, D. R., Keenan, F. P. 2004, ApJ, 604, 936
\bibitem{} Bogdan, T. J.,  Carlsson, M. Hansteen, V., McMurry, A., 
Rosenthal, C. S.,  
Johnson, M., Petty-Powell, S., Zita, E. J.,Stein, R. F., McIntosh, S. W., \&
Nordlund, \AA  2003, ApJ, 599, 626
\bibitem{} Campos, L. M. B. C. 1987, Rev. Mod. Phys. 59, 363
\bibitem{} Cauzzi, G., Falchi, A., \& Falciani, R. 2000, A\& A, 357, 1093
\bibitem{} Choudhuri, A. R., Auffret, H., \& Priest, E. R. 1993, Sol. Phys., 143, 49
\bibitem{} Curdt, W., \& Heinzel, P. 1998, ApJ, 503, L95
\bibitem{} Hasan, S. S., \& Kalkofen, W. 1999, ApJ, 519, 899
\bibitem{} Hasan, S. S., Kalkofen, W., \& van Ballegooijen, A. A. 2000, ApJ, 535, L67
\bibitem{} Hasan, S. S., Kalkofen, W., van Ballegooijen, A. A., \& Ulmschneider,
P. 2003, ApJ 585, 1138 
\bibitem{} Heinzel, P., \& Curdt, W. 1999, in Third Advances in Solar Physics
Euroconference: Magnetic Fields and Oscillations, eds. B. Schmieder,
A. Hofmann, \& J. Staude, ASP Conf. Ser., Vol. 184, 201
\bibitem{} Hollweg, J. V., Jackson, S., \& Galloway, D. 1982, Sol. Phys., 75, 35
\bibitem{} Huang, P., Musielak, Z. E., \& Ulmschneider, P. 1995, A \& A, 297, 579
\bibitem{} Judge, P., Carlsson, M., \& Wilhelm, K. 1997, ApJ, 490, L195
%\bibitem{} Judge, P., Tarbell, T. D., \& Wilhelm, K. 2001, ApJ, 554, 424
\bibitem{} Kalkofen, W. 1997, ApJ., 486, L148
\bibitem{} Lites, B. W., Rutten, R. J., \& Kalkofen, W. 1993, ApJ, 414, 345
\bibitem{}
McAteer, R. T. J., Gallagher, P. T., Williams, D. R., Mathioudakis,
M., Phillips, K. J. H. \& Keenan, F. P. 2002, ApJ, 567, L168
\bibitem{}
McAteer, R. T. J., Gallagher, P. T., Williams, D. R., Mathioudakis,
M., Phillips, K. J. H. \& Keenan, F. P. 2002, ApJ, 567, L168
%\bibitem{} McIntosh, S. W., Bogdan, T. J., Cally, P. S., Carlsson, M., Hansteen,
%V. H., Judge, P. G., Lites, B. W., Peter, H., Rosenthal, C. S., \&
%Tarbell, T. D. 2001, ApJ, 548, L237
Muller, R. 1983, Solar Phys, 85, 113
Muller, R. 1985, Solar Phys, 100, 237
%\bibitem{} Muller, R., Roudier, Th., Vigneau, J., \& Auffret, H.\ 1994, A \& A,
283, 232
\bibitem{} Musielak, Z. E., Rosner, R., \& Ulmschneider, P. 1989, ApJ, 337, 470
\bibitem{} Musielak, Z. E., Rosner, R., Gail, H. P., \& Ulmschneider, P. 1995,
ApJ, 448, 865 
\bibitem{} Musielak Z.E., \& Ulmschneider P.  2001, A\&A, 370, 541
\bibitem{} Musielak Z.E., \& Ulmschneider P.  2002, A\&A, 386, 606
\bibitem{} Musielak Z.E., \& Ulmschneider P.  2003a, A\&A, 400, 1057
\bibitem{} Musielak Z.E., \& Ulmschneider P.  2003b, A\&A, 406, 725
\bibitem{} Noble M.W., Musielak Z.E., \& Ulmschneider P. 2003, A\&A, 409, 1085
\bibitem{} Osin, A., Volin, S., \& Ulmschneider, P. 1999, A \& A, 351, 359
\bibitem{} Roberts, B., \& Ulmschneider, P. 1997, in Solar and Heliospheric Plasma Physics,
           Simett G.M., Alissandrakis C.E., Vlahos L. Eds., Lecture Notes in Physics 489,
           Springer Verlag, Heidelberg, Berlin, p. 75
\bibitem{}
Rosenthal, C. S., Bogdan, T. J., Carlsson, M., Dorch, S. B. F.,
Hansteen, V., McIntosh, S. W., McMurry, A., Nordlund, \AA \& Stein,
R. F. 2002, ApJ, 564, 508
%\bibitem{} Ryutov, D. D., \& Ryutova, M. P. 1976, Sov. Phys. J.E.T.P., 43, 491
\bibitem{} Spruit, H. C. 1982, Sol. Phys., 75, 3
\bibitem{} Ulmschneider, P. 2003, in Lectures on Solar Physics, Antia H.M., Bhatnagar A., 
           Ulmschneider P., Eds., Lecture Notes in Physics 619, 
           Springer Verlag, Heidelberg, Berlin, p. 232 
\bibitem{} Ulmschneider, P., \& Musielak, Z. E. 1998, A \& A, 338, 311
\bibitem{} Ulmschneider, P., Nowak, T., Bohn, U., \& Kalkofen, W. 1977, A \& A,
54, 61
\bibitem{} Ulmschneider, P., Z\"ahringer, K.,  \& Musielak, Z. E. 1991, A \& A,
241, 625
%\bibitem{} von Uexk\"ull, M., \& Kneer, F.\ 1995, A \& A, 294, 252
\bibitem{} Zhugzhda, Y.\ D., Bromm, V., \& Ulmschneider, P.\ 1995, A \& A, 300, 302
\end{thebibliography}
\end{document}